\documentclass[11pt,preprint,a4]{aastex}

\slugcomment{To appear in the Astrophysical Journal}

\begin{document}

\title{An Improved Virial Estimate of Solar Active Region Energy}
\author{M. S. Wheatland}
\affil{School of Physics, University of Sydney, \\
NSW 2006, Australia} 
\email{m.wheatland@physics.usyd.edu.au}
\author{Thomas R. Metcalf}
\affil{Colorado Research Associates Division, Northwest Research 
Associates, Inc., \\
3380 Mitchell Ln., Boulder, CO 80301} 
\email{metcalf@cora.nwra.com}

\begin{abstract}
The MHD virial theorem may be used to estimate the magnetic energy of 
active regions based on vector magnetic fields measured at the photosphere 
or chromosphere. However, the virial estimate depends on the measured 
vector magnetic field being force-free. Departure from force-freeness 
leads to an unknown systematic error in the virial energy estimate, and 
an origin dependence of the result. We present a method for estimating 
the systematic error by assuming that magnetic forces are
confined to a thin layer near the photosphere. If vector magnetic field
measurements are available at two levels in the low atmosphere 
(e.g.\ the photosphere and the chromosphere), the systematic error may be 
directly calculated using the observed horizontal and vertical field 
gradients, resulting in an energy estimate which is independent of the 
choice of origin. If (as is generally the case) measurements are available 
at only one level, the systematic error may be approximated using the 
observed horizontal field gradients together with a simple linear 
force-free model for the vertical field gradients. The resulting 
`improved' virial energy estimate is independent of the choice of 
origin, but depends on the choice of the model for the vertical field 
gradients, i.e.\ the value of the linear force-free parameter $\alpha$. 
This procedure is demonstrated for five vector magnetograms, including 
a chromospheric magnetogram.
\end{abstract}


\section{Introduction}

There is considerable interest in the possibility of estimating energies
for solar active regions from photospheric or chromospheric vector
magnetic field measurements. Solar flares derive their energy from 
magnetic fields in active regions, and it is possible that reliable 
estimates of active region energy could be used to improve the accuracy 
of solar flare prediction. 
The problem is timely given the wealth of magnetic field measurements 
shortly to be available from the next generation of ground- and 
space-based vector magnetographs, in particular SOLIS~\citep{kel&03}, 
and upcoming Solar Dynamics Observatory and Solar-B instruments.

One approach to the problem involves assuming the
field is force-free, i.e.\ has a zero Lorentz force density 
${\bf f}=\mu_0^{-1}(\nabla\times {\bf B})\times {\bf B}$ 
at and above the level of the measurements. In that case the simple 
form of the MHD virial theorem 
(Chandrasekhar 1961, chap.\ 13; Molodensky 1974)
\begin{equation}\label{eq:simple_virial}
\frac{1}{2\mu_0}\int_{z>0}B^2dV=
\frac{1}{\mu_0}\int_{z=0} (B_x x+B_y y) B_z\,dxdy
\end{equation}
applies. In this equation $z=0$ denotes the photosphere or 
chromosphere, which is approximated by a plane. In application to solar 
active regions, the interest is with the magnetic free energy, i.e.\ 
the energy associated with electric currents in the corona, rather than 
the total magnetic energy. This may be calculated by subtracting off the 
virial estimate $U$ for the potential field with the same $B_z$ 
distribution at $z=0$ (Low 1982). 

There are many difficulties associated with applying 
Equation~(\ref{eq:simple_virial}) to vector magnetic field measurements
in solar active regions. The measured field values are noisy, 
in particular the transverse (to the line of sight) field values, 
and the errors may compromise the virial energy estimate 
(Klimchuk, Canfield \& Rhoads 1992; Wheatland 2001). 
Error is also introduced by faults in the resolution of the 180 degree
ambiguity in the direction of the transverse field (e.g.\ Metcalf 1994). 
A potentially more serious problem is that the magnetic field is not 
strictly force-free at the level of the photosphere, where most vector 
magnetic field measurements originate, so the basic assumption 
underlying Equation~(\ref{eq:simple_virial}) is not met.  A related 
point is that the right hand side of Equation~(\ref{eq:simple_virial}) 
is dependent on the choice of the origin of the co-ordinate system, 
unless the net horizontal Lorentz force on the volume vanishes. 
Specifically, if we denote the integral on the right hand side of 
Equation~(\ref{eq:simple_virial}) by $W_0$, then if the origin is 
changed to $(x_0,y_0)$, the integral becomes
\begin{equation}\label{eq:W0p}
W_0^{\prime}=W_0+x_0F_x+y_0F_y,
\end{equation}
where 
\begin{equation}\label{eq:Fx}
F_x
=\int_{z>0}f_x\,dV
=-\frac{1}{\mu_0}\int_{z=0} B_xB_z\,dxdy 
\end{equation}
and
\begin{equation}\label{eq:Fy}
F_y
=\int_{z>0}f_y\,dV
=-\frac{1}{\mu_0}\int_{z=0} B_yB_z\,dxdy
\end{equation}
are the net horizontal Lorentz forces on the volume (Molodenskii 1969;
Aly 1984). Hence for non-zero $F_x$ or $F_y$ there is no unique virial 
energy estimate.

Equation~(\ref{eq:simple_virial}) was first applied to photospheric vector 
magnetic field data by Gary, Moore \& Hagyard (1987) and Sakurai (1987). 
Gary, Moore \& Hagyard (1987) found $W_0-U$ to be of order 
$10^{25}\,{\rm J}$ for
one active region. Although they did not state the choice of origin 
for this result, they investigated the influence of random errors in 
individual field measurements on the potential component of the energy, 
and concluded that the non-potential component was $3\sigma$
above the noise. Sakurai (1987) was less successful: for 120 vector
magnetograms he found $(W_0-U)/U\approx 0.085\pm 0.81$. 
However, Sakurai did not mention the choice of origin. Klimchuk, 
Canfield \& Rhoads (1992) investigated the influence of polarisation 
measurement errors on energy estimates made using 
Equation~(\ref{eq:simple_virial}), and concluded that the
detection of non-potential energy may be possible for the 
most energetic active regions. However, they also did not consider the 
problem of the origin-dependence of the estimates. 

Metcalf et al.\ (1995) appear to be the first authors to explicitly
consider the consequence of non-zero Lorentz forces for the practical
application of the virial theorem. They examined the height dependence 
of net Lorentz forces for an active region by making observations 
using the Mees Solar Observatory Stokes Polarimeter at different offsets 
from the center of the chromospheric Na I 
$\lambda$5896 line, thereby sampling different heights 
in the atmosphere. They found that the forces become small about 
500~km above the photosphere. They applied 
Equation~(\ref{eq:simple_virial}) to the field determined at each 
height, and averaged over a number of choices of origin
within the field of view. McClymont, Jiao \& Mikic (1997) subsequently
pointed out that the virial energy estimate~(\ref{eq:simple_virial}) 
is dimensionally of order $B^2L^3/\mu_0$, where $B$ is the average field 
strength and $L$ is the horizontal extent of the active region, 
whereas the true energy is of order $B^2L^2H/\mu_0$, where $H< L$ is 
the scale height of the field. This implies that there is substantial
cancellation in the virial integral. It follows from 
Equation~(\ref{eq:W0p}) that if $F_iL/(B^2L^3/\mu_0)$ is of 
order $H/L$ (for $i=x,y$), then the change in the virial energy 
associated with moving the origin across the field of view is comparable 
to the true energy. 
McClymont, Jiao \& Mikic (1997) gave an example of applying 
Equation~(\ref{eq:simple_virial}) to a particular active region,
and by varying the origin over the field of view found total energy 
estimates ranging from $1.23\times 10^{26}\,{\rm J}$ to $1.92\times
10^{26}\,{\rm J}$, compared with the (unique) potential field energy
$1.65\times 10^{26}\,{\rm J}$. They concluded that ``any virial
theorem estimate of the free energy is meaningless'', and suggested that
it is better to calculate a three dimensional
magnetic field from boundary values and estimate its energy. However,
calculating a realistic model of the field (e.g.\ a nonlinear force-free 
model) based on observed boundary values is itself a challenging 
problem. Other authors have also estimated net Lorentz forces 
from vector magnetic field data. Based on 12 photospheric vector 
magnetograms, Moon et al.\ (2002) reported downward Lorentz forces of 
order $0.06-0.32$ times the magnetic pressure force, 
with a median value of 0.13. More recently Georgoulis \& LaBonte (2004)
estimated net vertical Lorentz forces for a number of photospheric
vector magnetograms using a novel technique and reported large values.

The Mees Solar Observatory Imaging Vector Magnetograph is now routinely
producing chromospheric magnetograms using the Na I $\lambda$5896 line, 
which as noted above is formed at a height in the atmosphere which
is close to being force free (Metcalf et al.\ 1995). 
Recently Metcalf, Leka \& Mickey
(2005) calculated total and free magnetic energies for active region
10486 (which produced a number of the large `Halloween flares' of
October-November 2003) using chromospheric vector magnetograms from
Mees. They were unable to identify a change in energy associated with the 
X10 flare of 29 October 2003, but they determined the free energy of the
active region at three different times to be of order 
$10^{26}\,{\rm J}$. They also found the field to be almost force free, 
with ratios of the net horizontal Lorentz forces to 
the net magnetic pressure force of order $0.01$.
Following Metcalf et al.\ (1995), Metcalf, Leka \& Mickey applied
Equation~(\ref{eq:simple_virial}) with a `pseudo-Monte Carlo' procedure 
involving choosing many different origins within the field of view and
averaging the results. Their quoted errors are standard deviations
corresponding to this procedure. 

In this paper we identify the source of the origin dependence of 
virial energy estimates of the total magnetic energy as the neglect of 
additional terms involving volume integrals of spatial moments of the 
Lorentz force density. We point out that the missing terms may be 
calculated in straightforward ways assuming (as is observed) that the 
Lorentz forces are confined to a thin layer above the photosphere. If 
measurements of the vector magnetic field are available at two heights 
in the low atmosphere, e.g.\ at the photosphere and chromosphere, then 
the missing terms may be calculated directly using observed horizontal 
and vertical field gradients. It is more common to have measurements 
at a single height, for example from a photospheric vector magnetograph. 
In this case the missing terms may be approximated using the observed 
horizontal field gradients together with vertical field gradients 
from a simple linear force-free model. The result is 
termed an `improved' virial energy estimate. By construction it is 
independent of the choice of origin, although it depends on the choice 
of the model for the vertical field gradients. We also investigate the 
pseudo-Monte Carlo method of averaging Equation~(\ref{eq:simple_virial}) 
over choices of origin within the field of view. We show that this method 
is expected to give reliable estimates of the magnetic energy if the 
Lorentz forces are approximately uniformly distributed across the region 
of interest. More generally the distribution of Lorentz forces needs to 
be modelled, for example using the methods outlined here. We further 
show that the pseudo-Monte Carlo procedure is equivalent to the simple 
virial estimate with the origin at the center of the field of view, 
and we obtain an analytic expression for the standard deviation associated 
with the averaging procedure. The improved virial method is then 
demonstrated by application to five sets of vector magnetic field data, 
including a chromospheric vector magnetogram.

The layout of the paper is as follows. In section~2 the origin
dependence of the virial integral is discussed, and the improved virial
method is outlined. In section~3 the improved method is applied to 
vector magnetograms, and in section~4 the results are discussed. The
Appendix presents analytic results corresponding to the pseudo-Monte 
Carlo procedure.
  
\section{Method}

\subsection{The origin dependence of the virial integral}

A form of the virial theorem taking into account the possibility of 
non-zero Lorentz forces is (Molodenskii 1969)
\begin{equation}\label{eq:virial_hs}
\frac{1}{2\mu_0}\int_{z>0} B^2 \,dV
=\frac{1}{\mu_0}\int_{z=0} (B_x x+B_y y) B_z\,dxdy
+\int_{z>0}{\bf r}\cdot {\bf f}\,dV,
\end{equation}
where ${\bf f}$ is the Lorentz force density. 
This expression may be rewritten as 
\begin{equation}\label{eq:W=W0+W1}
W=W_0+W_1,
\end{equation}
where $W$ denotes the total magnetic energy, $W_0$ is the simple virial
estimate appearing on the right hand side of
Equation~(\ref{eq:simple_virial}), and $W_1$ is the final integral in
Equation~(\ref{eq:virial_hs}). This latter term may be written
\begin{equation}
W_1=\langle x\rangle F_x+\langle y\rangle F_y+\langle z \rangle F_z,
\end{equation}
where
\begin{equation}\label{eq:meanx}
\langle x\rangle = \left. \int_{z>0}f_xx \,dV\right/ \int_{z>0}f_x \,dV, 
\end{equation}
\begin{equation}\label{eq:meany}
\langle y\rangle = \left. \int_{z>0}f_yy \,dV\right/ \int_{z>0}f_y \,dV,
\end{equation}
and
\begin{equation}\label{eq:meanz}
\langle z\rangle = \left. \int_{z>0}f_zz \,dV\right/ \int_{z>0}f_z \,dV
\end{equation}
are Lorentz force-weighted average positions. There is observational
evidence that the Lorentz forces are confined to a narrow layer close to
the photosphere (Metcalf et al.\ 1995). Hence we make the approximations
\begin{eqnarray}
\int_{z>0}f_xx\,dV &\approx &\Delta z\int_{z=0}f_xx\,dxdy, \nonumber \\
\int_{z>0}f_x\,dV &\approx &\Delta z\int_{z=0}f_x\,dxdy,
\end{eqnarray}
where $\Delta z$ is the thickness of the forcing layer, and similarly
for the integrals in Equations~(\ref{eq:meany}) and~(\ref{eq:meanz}). 
This leads to
\begin{equation}\label{eq:meanx1}
\langle x \rangle \approx \left. \int_{z=0}f_xx \,dxdy\right/
  \int_{z=0}f_x \,dxdy,
\end{equation}
\begin{equation}\label{eq:meany1}
\langle y \rangle \approx \left. \int_{z=0}f_yy \,dxdy\right/
  \int_{z=0}f_y \,dxdy,
\end{equation}
and $\langle z\rangle =0$.

For non-zero Lorentz forces, $W_1$ will (in general) be non-zero. 
The term $W_1$ then represents a systematic error in the
simple estimate~(\ref{eq:simple_virial}) of the magnetic energy.
In general this term cannot be estimated precisely by considering how 
$W_0$ changes under various changes of origin, e.g.\ by moving the origin 
within the field of view (Metcalf et al.\ 1995; Metcalf, Leka \& 
Mickey 2005). From Equation~(\ref{eq:W0p}), the change in $W_0$ 
due to shifting the origin depends only on the net Lorentz force 
components $F_x$ and $F_y$, which may be estimated using 
Equations~(\ref{eq:Fx}) and~(\ref{eq:Fy}). However, the term 
$W_1$ depends on moments of the components of the Lorentz forces, 
which in general cannot be calculated from the net Lorentz 
forces alone. 

It is easy to see that by replacing $x$ and $y$ in 
Equation~(\ref{eq:meanx}) by $x-x_0$ and $y-y_0$, the terms 
$\langle x\rangle$ and $\langle y\rangle$ are replaced by 
\begin{equation}
\langle x\rangle^{\prime} = \langle x\rangle-x_0,
\quad {\rm and} \quad
\langle y\rangle^{\prime} = \langle y\rangle-y_0.
\end{equation}
Hence under this change of origin $W_1$ becomes
\begin{equation}
W^{\prime}_1=W_1-x_0F_x-y_0F_y.
\end{equation}
Combining this with Equation~(\ref{eq:W0p}), we see that $W=W_0+W_1$ 
is invariant under the change of origin, as required. In changing the
origin, both $W_0$ and $W_1$ change by the same amount, but the change 
gives no information about the true size of $W_1$.

It is straightforward to calculate the result of the pseudo-Monte
Carlo method of Metcalf, Leka \& Mickey (2005) in which $W_0$ is averaged
over all choices of origin within the field of view, subject to the
approximation that the measured magnetic field lies in a plane tangent
to the photosphere at the center of the field of view (see e.g.\ Gary \&
Hagyard 1990). If we denote the virial estimate with the origin 
corresponding to the lower left hand corner of the magnetogram by 
$W_0$, then from Equation~(\ref{eq:W0p}), the estimate with a choice of 
origin translated by $(x_0,y_0)$ in the heliographic plane is 
$W_0^{\prime}=W_0+x_0F_x+y_0F_y$. We require the average of
$W_0^{\prime}$ over all choices of $x_0$ and $y_0$ within the 
field of view, which may be written
\begin{equation}\label{eq:W0bar}
\overline{W_0^{\prime}}=W_0+\overline{x}_0F_x+\overline{y}_0F_y,
\end{equation}
where $\overline{x}_0$ denotes the average of $x_0$, and similarly for 
$\overline{y}_0$. In the Appendix it is shown that the average position
$(\overline{x}_0,\overline{y}_0)$ corresponds to the center of 
the image plane field of view. Comparing Equations~(\ref{eq:W0p}) 
and~(\ref{eq:W0bar}) it follows that $\overline{W_0^{\prime}}$ is
equal to the virial integral with the origin at the center of the
image plane field of view. The Appendix provides analytic expressions
for $\overline{x}_0$ and $\overline{y}_0$, as well as an expression
for the standard deviation associated with averaging $W_0^{\prime}$ 
over all choices of origin in the field of view, in terms of 
$F_x$, $F_y$, and the geometry. 

From Equations~(\ref{eq:meanx1}), (\ref{eq:meany1}), and~(\ref{eq:x0bar}) 
it follows that if $f_x$ and $f_y$ are constant in the plane $z=0$, we have 
$\langle x\rangle=\overline{x}_0$ and $\langle y \rangle=\overline{y}_0$. In
other words if the Lorentz forces are uniform, the Lorentz-force weighted 
average positions coincide with the average positions. In this case 
$W_1=\overline{x}_0F_x+\overline{y}_0F_y$, so
$W=W_0+\overline{x}_0F_x+\overline{y}_0F_y=\overline{W_0^{\prime}}$, 
i.e.\ the pseudo-Monte Carlo
method gives the correct answer. Somewhat more generally, if the 
Lorentz forces are approximately uniformly distributed over the region 
of interest, the simple form of the virial theorem with the origin at 
the center of the field of view gives approximately the correct answer 
for the total energy.

It is more realistic to assume that the forces are non-uniformly 
distributed, and in that case $\overline{W_0^{\prime}}$ will not give the 
correct energy. In the next section we consider how to model the 
Lorentz forces $f_x$ and $f_y$ over the field of view to give a more 
accurate, and origin-independent, estimate of the total magnetic energy.

\subsection{Estimating $f_x$ and $f_y$ in the plane $z=0$}

The basic idea is to approximate $f_x$ and $f_y$ in the plane $z=0$ 
based on available field gradients, and modelling where necessary. If 
the approximate versions of $f_x$ and $f_y$ are denoted
$\widetilde{f}_x$ and $\widetilde{f}_y$, then we can write for our 
estimates of the force-weighted positions 
\begin{equation}\label{eq:meanx2}
\widetilde{\langle x \rangle} = \left. \int_{z=0}\widetilde{f}_xx 
  \,dxdy\right/
  \int_{z=0}\widetilde{f}_x\, dxdy
\end{equation}
and
\begin{equation}\label{eq:meany2}
\widetilde{\langle y \rangle} = \left. \int_{z=0}\widetilde{f}_yy 
  \,dxdy\right/
  \int_{z=0}\widetilde{f}_y\, dxdy,
\end{equation}
and our estimate of the total energy becomes
\begin{equation}\label{eq:West}
\widetilde{W}= W_0+\widetilde{\langle x\rangle} F_x 
  +\widetilde{\langle y\rangle} F_y.
\end{equation}
We note in passing that the definitions Equations~(\ref{eq:meanx2}) 
and~(\ref{eq:meany2}) of the force-weighted positions require the 
integrals in the denominators to be non-zero.

The reason for writing the energy in this way is that 
$\widetilde{\langle x\rangle}$
and $\widetilde{\langle y\rangle}$ have the same behavior under a change 
of origin as the exact expressions 
$\langle x\rangle$ and $\langle y \rangle$. Specifically,
if the origin is changed to
$(x_0,y_0)$, then from Equations~(\ref{eq:meanx2}) and~(\ref{eq:meany2})
the force-weighted positions change to 
\begin{equation}
\widetilde{\langle x\rangle}^{\prime} = \widetilde{\langle x\rangle}-x_0,
\quad {\rm and} \quad
\widetilde{\langle y\rangle}^{\prime} = \widetilde{\langle y\rangle}-y_0.
\end{equation}
Combining this with Equations~(\ref{eq:W0p}) 
and~(\ref{eq:West}), it follows that $\widetilde{W}$ is 
{\em independent of the choice of origin}. Hence our procedure ensures 
an estimate of the energy which is
origin independent, a result which holds for any choices of 
$\widetilde{f}_x$ and $\widetilde{f}_y$. Of course, the value of 
$\widetilde{W}$ will depend on the specific choices. Hence we have 
gained origin independence, i.e.\ a unique value, at the expense of 
model dependence.

The Lorentz force density components $f_x$ and $f_y$ may be written
from their definition as
\begin{equation}\label{eq:f_x}
f_x=\frac{1}{\mu_0}
  \left[
  B_z\left(\frac{\partial B_x}{\partial z}
  -\frac{\partial B_z}{\partial x}\right)
  -B_y\left(\frac{\partial B_y}{\partial x}
  -\frac{\partial B_x}{\partial y}\right)
  \right]
\end{equation}
and 
\begin{equation}\label{eq:f_y}
f_y=\frac{1}{\mu_0}
  \left[
  B_x\left(\frac{\partial B_y}{\partial x}
  -\frac{\partial B_x}{\partial y}\right)
  -B_z\left(\frac{\partial B_z}{\partial y}
  -\frac{\partial B_y}{\partial z}\right)
  \right].
\end{equation}

The horizontal derivatives in these expressions may be estimated
from vector magnetic field data. If simultaneous measurements are 
available at two heights in the low atmosphere (e.g.\ at the photosphere 
and chromosphere), then the vertical derivatives may also be estimated 
from the data. However, measurements typically are available at only one
height. In that case the vertical derivatives are not specified by the 
data, and we adopt a procedure of modelling the derivatives based on
available information. Specifically, we use vertical derivatives from
a linear force-free model for the field. (Note that if the horizontal 
components of the linear force-free field match the observed $B_x$ and 
$B_y$ everywhere then $f_x=f_y=0$.) This procedure is similar to a 
common practice in the resolution of the 180 degree ambiguity, where the 
value of ${\rm div}\,{\bf B}$ is estimated from the data locally using linear 
force-free estimates of vertical gradients (e.g.\ Metcalf 1994).

The linear force-free model is defined by 
${\rm curl}\, {\bf B}=\alpha {\bf B}$, where $\alpha$ is the force-free 
constant. The $x$- and $y$-components of this equation give the 
expressions for the vertical gradients
\begin{equation}\label{eq:lfff_ddz}
\frac{\partial B_y}{\partial z}= -\alpha B_x +\frac{\partial
B_z}{\partial y}
\quad {\rm and} \quad
\frac{\partial B_x}{\partial z}= \alpha B_y + \frac{\partial B_z}{\partial x}.
\end{equation}
Combining Equations~(\ref{eq:f_x}), (\ref{eq:f_y}) and (\ref{eq:lfff_ddz})
gives the model forms for $f_x$ and $f_y$:
\begin{equation}\label{eq:ftilde_x}
\widetilde{f}_x(\alpha ) = \frac{1}{\mu_0}B_y
  \left(\alpha B_z-\mu_0J_z\right)
\end{equation}
and
\begin{equation}\label{eq:ftilde_y}
\widetilde{f}_y(\alpha ) = \frac{1}{\mu_0}B_x
  \left(\mu_0J_z-\alpha B_z\right),
\end{equation}
where
\begin{equation}
J_z=\frac{1}{\mu_0}\left(\frac{\partial B_y}{\partial x}
-\frac{\partial B_x}{\partial y}\right)
\end{equation}
is the vertical current density. The form of these expressions makes
clear the point that the estimates of the horizontal Lorentz forces will
be zero if the observed horizontal gradients match those of a force-free 
field, i.e.\ if $J_z=\alpha B_z/\mu_0$ is satisfied everywhere. 

Equations~(\ref{eq:meanx2}), (\ref{eq:meany2}), and~(\ref{eq:West})
together with~(\ref{eq:ftilde_x}) and~(\ref{eq:ftilde_y}) provide 
an `improved' virial energy estimate $\widetilde{W}(\alpha)$, which 
is independent of the choice of origin, but which depends on the choice
of the parameter $\alpha$. In Section~3 this estimate is evaluated for 
five vector magnetograms.

There are several sources of uncertainty in the improved virial 
estimate. There is
uncertainty associated with the modeling (i.e.\ the use of linear
force-free fields to approximate the vertical gradients), and with the
approximations inherent in Equations~(\ref{eq:meanx1})
and~(\ref{eq:meany1}). These uncertainties are hard to quantify. There
is also error in the energy estimate due to errors in the vector
magnetic field measurements, and due to the uncertainty in the choice of
$\alpha$. In Section~3 we will attempt to quantify these latter 
uncertainties, to give some measure of how reliable the estimate is.

A practical point to note in applying these formulae is that the 
horizontal derivatives and integrals are with respect to heliographic 
co-ordinates $(x,y)$. For data from vector magnetograms, these 
points form a non-uniform, non-Cartesian grid specified 
by the projection of points in the magnetograph field of view onto 
the photosphere or chromosphere. The evaluation of derivatives and
integrals is simplified by working with co-ordinates in the image plane, 
which are uniformly spaced. In the calculations in this paper we follow 
this practice, using the approximation that heliographic co-ordinates lie
on a plane tangent to the solar surface at the center of the region of
interest. In that case there is a simple linear transformation between
the co-ordinate systems (e.g.\ Gary \& Hagyard 1990; Venkatakrishnan \&
Gary 1990). 

\section{Application to vector magnetograms}

The improved virial method was applied to five vector magnetograms from 
the Mees Solar Observatory Imaging Vector Magnetograph (IVM). Four are 
photospheric magnetograms: 
NOAA active region 9077 observed on 14 July 2000 at 16:33UT 
(this is the Bastille Day 2000 flare region, observed after the
flare occurred); NOAA AR 10306 observed on 13 March 2003 at 17:14UT; 
NOAA AR 10386 observed on 20 June 2003 at 16:38UT; and NOAA AR 10397 
observed on 1 July 2003 at 18:18UT. These four magnetograms provide a
typical selection of data. The fifth dataset is one of the 
chromospheric vector magnetograms previously studied by Metcalf, 
Leka \& Mickey (2005): NOAA AR 10486, observed on 29 October 2003 at 
18:46UT. In each case the data correspond to measurements at one height 
in the solar atmosphere, and so we follow the procedure of modelling vertical
derivatives using a linear force-free model.

The IVM has 512 by 512 pixels with 0.55 arc second size, resulting in 
a field of view of about 4.7 arc minutes. (The corresponding projected
distance at 1~AU is $L\approx 2.0\times 10^8\,{\rm
m}$.) Details of the processing of IVM photospheric data are given in 
LaBonte, Mickey \& Leka (1999), and further details of the chromospheric 
magnetogram are given in Metcalf, Leka \& Mickey (2005). In each case 
the resolution of the azimuthal ambiguity was performed using the 
`minimum energy' method (Metcalf 1994). 

For each magnetogram the following procedure was followed. First the
fractional flux balance $\phi$ (the ratio of signed to unsigned magnetic 
flux) was evaluated. Next the energy $U$ of a potential field with the given 
boundary values $B_z$ was determined from the virial theorem using the 
Fourier transform potential field solution (Alissandrakis 1981), with
the Fourier transforms being performed with respect to image co-ordinates 
(see Venkatakrishnan \& Gary 1989). The simple virial estimate $W_0$ of 
Equation~(\ref{eq:simple_virial}) was then evaluated, with the origin 
correponding to the lower left hand corner of the magnetogram 
(this choice of origin is taken as the default for all origin-dependent 
terms). The forces $F_x$ and $F_y$ were also evaluated, together with the 
reference magnetic pressure force (Aly 1989)
\begin{equation}
F_0\equiv\frac{1}{2\mu_0}\int_{z=0} (B_x^2+B_y^2+B_z^2)\,dxdy.
\end{equation}
Then the estimate 
$\overline{W}_0=W_0+\overline{x}_0F_x+\overline{y}_0F_y$ 
corresponding to the pseudo-Monte Carlo method of Metcalf, Leka \& Mickey
(1995) was calculated using Equation~(\ref{eq:xybar}).

Next the parameter $\alpha$ was determined by minimizing the least squares
difference between the observed horizontal field and the horizontal 
field associated with the linear force-free field (Leka \& Skumanich 1999). 
An uncertainty $\sigma_{\alpha}$ was
also calculated, based on the variation in the goodness of fit as a
function of $\alpha$.  The functions $\widetilde{f}_x(\alpha)$ and
$\widetilde{f}_y(\alpha)$ were then evaluated at each gridpoint based 
on estimates of the horizontal derivatives from differencing, with the 
differencing performed in image plane co-ordinates (Gary \& Hagyard 1990). 
The resulting values were subject to $3\times 3$ median smoothing to 
remove spikes. The force-weighted positions 
$\widetilde{\langle x\rangle}(\alpha )$ and 
$\widetilde{\langle y\rangle}(\alpha )$ 
were then evaluated, and the improved virial energy 
$\widetilde{W}(\alpha )$ and the corresponding estimate of the free 
energy $\widetilde{W}(\alpha )-U$ were calculated.

Finally two uncertainties were numerically evaluated.
The error in the energy estimate due to the the errors in the field
measurements was estimated by a Monte Carlo procedure involving 100
evaluations of the improved virial energy estimate, with the field 
values $B_x$, $B_y$, and $B_z$ subject to random Gaussian perturbations
at each evaluation according to the estimated errors in the measurements. 
The standard deviation of the 100 estimates was taken as the estimate of
the uncertainty due to the field measurements, and this estimate is
denoted $\sigma ({\bf B})$. The second uncertainty considered is the error in
the estimate due to the uncertainty $\sigma_\alpha$ in the value of the
force-free parameter $\alpha$. It is straightforward to propagate the
uncertainty in $\alpha$ through to an uncertainty in
$\widetilde{W}(\alpha)$ using Equations~(\ref{eq:meanx2}),
(\ref{eq:meany2}), (\ref{eq:West}), (\ref{eq:ftilde_x}) 
and~(\ref{eq:ftilde_y}), and the resulting estimate is denoted 
$\sigma (\alpha)$. The two uncertainties were not combined because 
they describe quite different things.

The results are presented in Table 1. For AR 9077, the improved virial
estimate $\widetilde{W}_0(\alpha)-U\approx -2.0 \times 10^{25}\,{\rm J}$ 
is negative, which is unphysical.
However, for this case the estimate 
$\overline{W}_0-U\approx -4.6\times 10^{25}\,{\rm J}$ --- corresponding to
the pseudo-Monte Carlo method --- is also negative. The estimated
uncertainties $\sigma (\alpha )$ and $\sigma ({\bf B})$ do not account 
for the negative value, and hence in this case a physically meaningful 
estimate for the free energy is not obtained. 
For AR 10306, the improved virial estimate
$\widetilde{W}_0(\alpha)-U\approx 5.8\times 10^{25}$ is slightly
smaller than the pseudo-Monte Carlo estimate. 
However, the estimated uncertainties are both comparable to the 
improved virial estimate.
For AR 10386, the improved virial estimate 
$\widetilde{W}_0(\alpha)-U\approx 2.3\times 10^{25}\,{\rm J}$ is 
somewhat smaller than the pseudo-Monte Carlo estimate, whilst 
$\sigma (\alpha )$ is larger than the improved virial 
estimate. For AR 10397, 
the improved virial estimate 
$\widetilde{W}_0(\alpha)-U\approx 4.5\times 10^{26}\,{\rm J}$ is very 
large, even larger than the pseudo-Monte Carlo estimate. In this case
the uncertainties are somewhat smaller than the improved virial 
estimate. For the chromospheric magnetogram (AR 10486) the improved 
virial estimate $\widetilde{W}_0(\alpha)-U\approx 2.6\times 10^{26}\,{\rm J}$ 
is very similar to the pseudo-Monte Carlo value, and the estimated
uncertainties are an order of magnitude smaller. These results suggest
that a meaningful energy estimate is being obtained in this case.

The results for the photospheric magnetograms are mixed. There is once
instance (AR 9077) of a negative free energy estimate, and this is for a
large, apparently non-potential active region. For all of the
photospheric magnetograms the estimated uncertainties are large, and in
particular the uncertainties $\sigma(\alpha)$ associated with the model for
the vertical gradients are larger than the uncertainties
$\sigma ({\bf B})$ associated with the field measurements. 
These results suggest that although the improved virial method provides 
a unique estimate for the free energy, it may not always provide a 
meaningful estimate. The exception may be for active regions with very 
large free energies, such as AR 10397. For the chromospheric 
magnetogram, the improved virial estimate agrees with the simpler
pseudo-Monte Carlo estimate.

Finally we note that the pseudo-Monte Carlo value calculated for the
chromospheric magnetogram of AR 10486 is slightly different from the 
value $3.2\pm 0.7\times 10^{26}\,{\rm J}$ quoted by 
Metcalf, Leka \& Mickey (2005) for this magnetogram. The exact
computational procedure followed by Metcalf, Leka \& Mickey takes 
into account the curvature of the Sun's surface, which is here 
neglected, and this accounts for the difference. 

It is interesting to examine the spatial distributions of horizontal
Lorentz force obtained using Equations~(\ref{eq:ftilde_x}) 
and~(\ref{eq:ftilde_y}).
Figure~1 shows results for the chromospheric magnetogram of AR 10486. 
The left panel shows the vertical component of the magnetic 
field in the heliographic plane, and the right panel shows the
normalised horizontal Lorentz force 
$\widetilde{f}_h(\alpha)/{\rm max}[B_z^2/(2\mu_0)]$, where
$\widetilde{f}_h(\alpha) = \left[\widetilde{f}_x(\alpha)^2
  +\widetilde{f}_y(\alpha)^2\right]^{1/2}$. 
The vertical bar on the far right shows the range of values of the
horizontal Lorentz force.  
The horizontal and vertical white lines in the right panel correspond 
to $\widetilde{\langle x\rangle}(\alpha)/L$ and
$\widetilde{\langle y\rangle}(\alpha)/L$ respectively. 
The force-weighted positions are
found to correspond approximately with the center of the large positive
polarity region just below the center of the magnetogram. 

\begin{figure}
\plotone{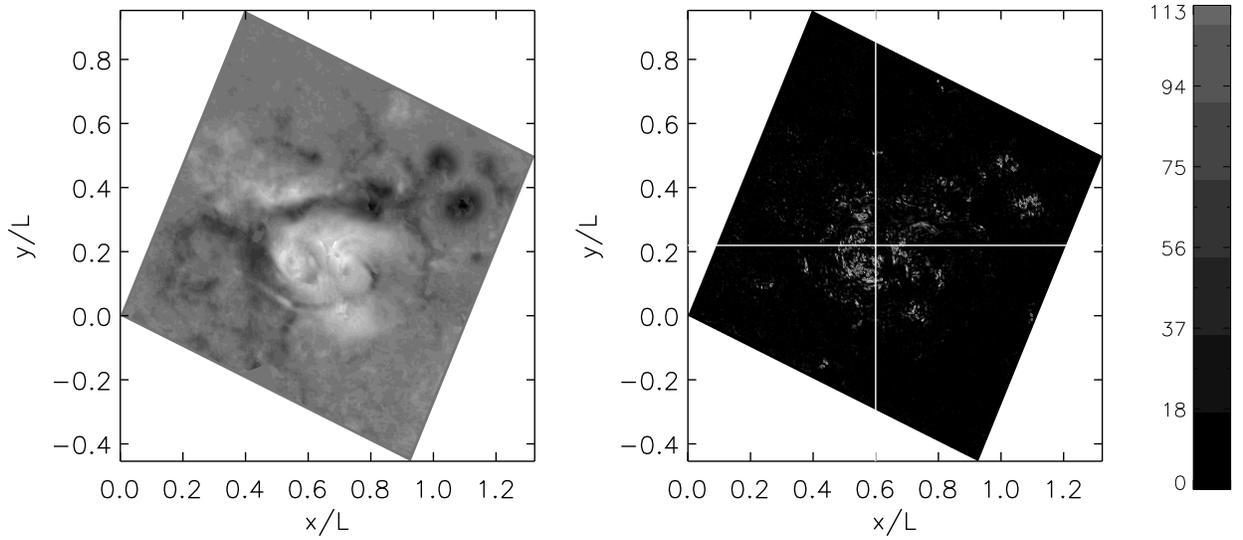}
\caption{Left panel: The vertical magnetic field from the chromospheric
magnetogram for AR 10486 at 18:46UT on 29 October 2003. 
Right panel: The 
estimate $\widetilde{f}_h(\alpha)/{\rm max}[B_z^2/(2\mu_0)]$ 
of the magnitude of the horizontal Lorentz force for the same 
magnetogram. The horizontal and vertical lines correspond 
to $\widetilde{\langle x\rangle}(\alpha)/L$ and 
$\widetilde{\langle y\rangle}(\alpha)/L$ respectively.} 
\end{figure}
 
\section{Discussion}

This paper addresses the problem of the origin dependence of virial
theorem estimates of the magnetic energy of solar active regions. A
method is presented for calculating an origin-independent energy based 
on estimation of terms missing from the usual statement of the virial
theorem, involving integrals of moments of the Lorentz forces. The method
assumes the forces are located close to the height where magnetic field
measurements are taken. If measurements are available at two heights in 
the low atmosphere, the terms may be directly estimated using observed
horizontal and vertical field gradients. If (as is generally the case)
measurements are available corresponding to a single height, the terms
may be estimated using observed horizontal field gradients together 
with vertical field gradients coprresponding to a linear force-free 
model of the field. This latter procedure is demonstrated in application 
to five vector magnetograms, including a chromospheric magnetogram, and 
the results are given in Table~1. The results suggest that, at least for 
photospheric magnetograms, the large uncertainties associated with field 
measurements and with the estimation of the force-free parameter makes 
obtaining a meaningful free energy estimate a difficult task. 

The accuracy of the pseudo-Monte Carlo method of Metcalf, Leka \&
Mickey (2005) is also investigated. It is shown that this method
provides a reliable estimate of magnetic energy provided Lorentz forces
are approximately uniformly distributed over the plane where measurements 
originate. More generally, it is necessary to take account of the 
distribution of Lorentz forces, for example using the methods described 
here. The result of the pseudo Monte-Carlo procedure is shown to 
correspond to the simple virial theorem estimate with the origin at the 
center of the field of view. An analytic expression for the standard 
deviation associated with the averaging procedure is also given 
[Equation~(\ref{eq:sigma0})].
 
The use of vertical gradients from a linear force-free field to estimate 
Lorentz forces is a crude approach, since linear force-free 
fields have identically zero Lorentz forces. However, the point is that 
the horizontal gradients of the observed field contain important 
information about the Lorentz forces. Although the 
vertical gradients are inaccurate, the improved virial estimate contains 
the additional information in the horizontal gradients, and hence may 
provide a more accurate estimate of the total energy. The approach is
similar to one already widely used in ambiguity resolution, in which 
observed horizontal field gradients together with the vertical gradients 
of a linear force-free field are used to estimate ${\rm div}\,{\bf B}$ 
(Metcalf 1994).
 
It should be noted that the method outlined here is more general than 
the specific approach using the linear force-free vertical gradients, 
and may be improved by using better estimates of the vertical gradients, 
if available. We note that the uncertainties $\sigma (\alpha)$ in the 
energy estimates associated with the linear force-free vertical gradients 
were found to be larger than the uncertainties $\sigma ({\bf B})$ associated 
with the field measurements. This also suggests the need for improved 
estimates of the vertical field gradients. It is expected that the method 
will provide better estimates if magnetograms giving the field 
simultaneously at two heights are available (e.g.\ at the photospheric 
and chromospheric levels), and we are currently investigating this. 
Such a procedure will also allow a test of the use of the linear 
force-free model for the vertical gradients. 

The nature of the non-zero Lorentz forces has not been described. In
general there may be local pressure, gravity, and inertial forces
present at the level of the measurements, which will lead to local
non-zero values of $f_x$ and $f_y$. It should also be noted that there
may be substantial non-zero net Lorentz forces due to flux closing
outside the field of view of a magnetogram, even if the local Lorentz
forces are themselves small. Generally magnetograms only have flux
balanced to within 10\% or so, since fields of view are relatively
small, and active regions can have connections to surrounding flux 
(e.g.\ see the top row of Table~1), so this is an important 
consideration. This effect will also be present with chromospheric
magnetograms. We note that the present method is able to model 
this kind of global departure from force-freeness.   

With the advent of improved vector magnetographs, including instruments
in space, it is important that as much information as possible is
extracted from the data, and the present method may be useful in this 
regard. Hopefully this work goes some way towards countering the 
assertion of McClymont, Jiao \& Mikic (1997) that ``any virial theorem 
estimate of the free energy is meaningless''.

\acknowledgments
We acknowledge that the use of the chromospheric magnetic field data was
made possible with funding from NASA NAG5-12466 and AFOSR 
F49620-03-C-0019, and we thank K.D. Leka for her work on this data. 
Paul Watson and Alex Judge provided helpful discussions. We also 
acknowledge the support of an Australian Research Council QEII 
Fellowship, and thank the referee for a number of constructive comments.

\appendix
 
\section{Averaging $W_0$ over origins within the field of view}

The pseudo-Monte Carlo method of Metcalf, Leka \& Mickey (2005)
involves averaging the simple virial estimate~(\ref{eq:simple_virial})
over many choices of origin within the field of view. 
It is straightforward to calculate the resulting average and
standard deviation based on a single evaluation
of the virial integral together with $F_x$ and $F_y$, subject to the
assumption that the measured field lies in a plane tangent to the 
photosphere. 
Denoting the virial estimate with the origin corresponding to the 
lower left hand corner of the magnetogram by $W_0$, then following 
Equation~(\ref{eq:W0p}), the estimate with a choice of origin
$(x_0,y_0)$ is $W_0^{\prime}=W_0+x_0F_x+y_0F_y$. We require the average of 
$W_0^{\prime}$ over choices of $x_0$ and $y_0$, which we denote
\begin{equation}\label{eq:pseudo_analytic}
\overline{W_0^{\prime}}=W_0+\overline{x}_0F_y+\overline{y}_0F_y,
\end{equation}
where
\begin{equation}\label{eq:x0bar}
\overline{x}_0=\left.\int_{z=0}x_0dx_0dy_0\right/ \int_{z=0}dx_0 dy_0,
\end{equation}
and similarly for $\overline{y}_0$.

The integrals for $\overline{x}_0$ and $\overline{y}_0$ may be evaluated 
by changing variables in the 
integration to image plane co-ordinates $\xi$ and $\eta$, aligned with 
terrestrial north and west respectively, and integrating over 
$0\leq \xi\leq L$, $0\leq \eta \leq L$. The image plane and heliographic 
co-ordinates are related by the linear transformation 
(e.g.\ Gary \& Hagyard 1990)
\begin{equation}\label{eq:linear_trans}
\left(\begin{array}{c} 
      \xi \\ \eta
      \end{array}
\right)
=
\left(\begin{array}{cc} 
      c_{11} & c_{12} \\ c_{21} & c_{22}
      \end{array}
\right)
\left(\begin{array}{c} 
      x \\ y 
      \end{array}
\right),
\end{equation} 
where
\begin{eqnarray}
c_{11} &=& \cos P\cos \ell_c-\sin P\sin B_0\sin \ell_c, \nonumber \\
c_{12} &=& -\cos P\sin B_c\sin \ell_c-\sin P (
  \cos B_0\cos B_c + \sin B_0\sin B_c\cos \ell_c), \nonumber \\
c_{21} &=& \sin P\cos \ell_c+\cos P\sin B_0\sin \ell_c, \nonumber \\
c_{22} &=& -\sin P\sin B_c\sin \ell_c+\cos P (
  \cos B_0\cos B_c + \sin B_0\sin B_c\cos \ell_c).
\end{eqnarray}
In these expressions $B_0$ is the latitude of the disk center and $P$ is
the angle between the projection of the rotation axis of the Sun on the
sky and terrestrial north, measured eastward from north. The angle
$\ell_c$ is the longitude of the center of the region of interest with 
respect to the meridian passing through disk center (the Central
Meridional Distance, or CMD), and $B_c$ is the latitude of the center of
the region of interest. 

To perform the integrations it is also necessary to calculate the Jacobian
of the transformation~(\ref{eq:linear_trans}), which is the reciprocal of 
the cosine of the angle $\Theta$ between the 
normal vector to the tangent plane and the line-of-sight to the observer. 
Specifically we require
\begin{eqnarray}\label{eq:jacobian}
\frac{\partial (x,y)}{\partial (\xi ,\eta )}&=&
\left|\begin{array}{cc}
      c_{11} & c_{12} \\ c_{21} & c_{22}
      \end{array}
\right|^{-1} \nonumber \\
&=&
\left( \sin B_0\sin B_c +\cos B_0\cos B_c \cos \ell_c \right)^{-1} 
  \nonumber \\
&=&\sec\Theta .
\end{eqnarray}

Using Equations~(\ref{eq:linear_trans}) and~(\ref{eq:jacobian}), 
the integrals for $\overline{x}_0$ and $\overline{y}_0$ evaluate to
\begin{eqnarray}\label{eq:xybar}
\overline{x}_0&=&\sec\Theta\frac{L}{2}(c_{22}-c_{12}), \nonumber \\
\overline{y}_0&=&\sec\Theta\frac{L}{2}(-c_{21}+c_{11}).
\end{eqnarray}
From the inverse of the transformation~(\ref{eq:linear_trans}) it 
follows that $(\overline{x}_0,\overline{y}_0)$ corresponds to
the center of the image plane field of view:
\begin{eqnarray}\label{eq:xybar2}
\overline{x}_0 &=& x(\xi=L/2,\eta=L/2), \nonumber \\
\overline{y}_0 &=& y(\xi=L/2,\eta=L/2).
\end{eqnarray}
Hence the estimate $\overline{W_0^{\prime}}$ corresponds to choosing the 
origin at the center of the field of view.

Following a similar procedure the standard deviation 
\begin{equation}
\sigma_0=\left\{
  \overline{(W_0^{\prime}-\overline{W_0^{\prime}})^2}\right\}^{1/2}
\end{equation}
may be evaluated to give
\begin{equation}\label{eq:sigma0}
\sigma_0=\sec\Theta\frac{L}{2\sqrt{3}}\left[
(c_{22}^2+c_{12}^2)F_x^2-2F_xF_y(c_{22}c_{21}+c_{12}c_{11})
+(c_{21}^2+c_{11}^2)F_y^2\right]^{1/2}.
\end{equation}  

Finally, it should be noted that the computational procedure followed by
Metcalf, Leka \& Mickey (2005) takes into account the curvature of the
Sun's surface, which is neglected here. When curvature is included, the 
results of the pseudo-Monte Carlo procedure will be slightly different 
from the estimates given by Equations~(\ref{eq:pseudo_analytic}), 
(\ref{eq:xybar}), and~(\ref{eq:sigma0}).

\clearpage
\begin{table}
\begin{center}
\caption{Results of applying the method to five vector 
magnetograms.\label{tbl-1}}
\begin{tabular}{c|ccccc}
\tableline\tableline
&  AR 9077 & AR 10306 & AR 10386 &  AR 10397 & AR 10486 \\
\tableline
$\phi$ & -0.099 & 0.46 & -0.14 & 0.37 & -0.0013 \\ 
$U$ [J] & $5.9\times 10^{25}$ & $1.6\times 10^{26}$ 
  & $3.2\times 10^{25}$ & $1.5\times 10^{26}$ & $4.4\times 10^{26}$ \\ 
$W_0$ [J] & $-1.5\times 10^{26}$ & $5.5\times 10^{26}$ 
  & $-3.0\times 10^{26}$ & $-3.1\times 10^{26}$ & $7.7\times 10^{26}$ \\ 
$\overline{W}_0$ [J] & $5.4\times 10^{25}$ & $2.4\times 10^{26}$ 
  & $1.1\times 10^{26}$ & $2.4\times 10^{26}$ & $7.0\times 10^{26}$ \\ 
$\overline{W}_0-U$ [J] & $-4.6\times 10^{24}$ & $8.0\times 10^{25}$ 
  & $7.7\times 10^{25}$ & $9.9\times 10^{25}$ & $2.6\times 10^{26}$ \\ 
$|F_x/F_0|$ & 0.0041 & 0.0060 & 0.093 & 0.13 & 0.0055\\ 
$|F_y/F_0|$ & 0.079 & 0.046 & 0.075 & 0.027 & 0.0041\\ 
$\alpha \pm \sigma_{\alpha}$ [$L^{-1}$] 
  & $-2.8\pm 7.1$ & $0.2\pm 8.9$ & $15.1\pm 10.5$ & $-0.4\pm 8.4$ 
  & $-8.3\pm 11.2$\\ 
$\widetilde{\langle x\rangle}(\alpha)/L$ &  0.54 & 0.084 & 0.38 & 0.64 
  & 0.60 \\ 
$\widetilde{\langle y\rangle}(\alpha)/L$ & 0.44 & 0.70 & 0.53 & 0.14 
  & 0.22 \\ 
$\widetilde{W}(\alpha )-U$ [J] & $-2.0\times 10^{25}$ & $5.8\times 10^{25}$ 
  & $2.3\times 10^{25}$ & $4.5\times 10^{26}$ & $2.6\times 10^{26}$ \\ 
$\sigma ({\bf B})$  [J] & $4.2\times 10^{24}$ & $5.5\times 10^{24}$ 
  & $1.1\times 10^{24}$ & $9.0\times 10^{25}$ & $1.1\times 10^{25}$ \\
$\sigma (\alpha)$   [J] & $8.3\times 10^{24}$ & $3.0\times 10^{25}$ 
  & $6.2\times 10^{25}$ & $2.6\times 10^{26}$ & $1.2\times 10^{25}$ \\
\tableline
\end{tabular}

\end{center}
\end{table}

\end{document}